\documentclass{aa}

\usepackage{graphicx}
\usepackage{txfonts}
\usepackage{natbib}
\usepackage{amsmath}
\usepackage{bbold}
\usepackage{tikz}
\usepackage{multirow}
\usepackage{xspace}
\usepackage[colorlinks=true, linkcolor=blue, citecolor=blue, filecolor=blue, urlcolor=blue]{hyperref}
\usepackage{placeins}

\def\T{^\mathrm{T}}
\def\d{\mathrm{d}}

\bibpunct{(}{)}{;}{a}{}{,} 
\long\def\orcidlink#1{\unskip\protect\href{https://orcid.org/#1}{\protect\includegraphics[width=8pt,clip]{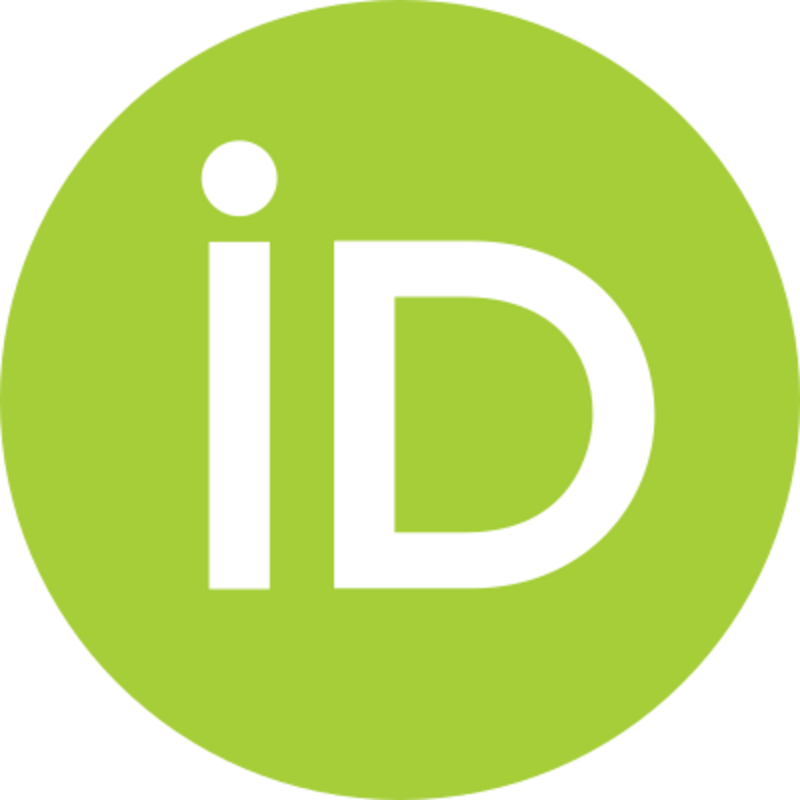}}}

\begin{document}

\title{Unveiling the mutual inclination distribution in astrometric surveys}

\author{J.-B.~Delisle \inst{\orcidlink{0000-0001-5844-9888}} \corrauth{jean-baptiste.delisle@unige.ch}}

\institute{Département d'astronomie, Université de Genève, chemin Pegasi 51, 1290 Versoix, Switzerland}

\date{Received 19 August 2026 / Accepted 29 October 2026}

\abstract{
  Mutual inclination in three (or more) body systems (triple stars, circumbinary planets, multi-planetary systems, etc.)
  is a tracer of their formation and evolution.
  Gaia Data Release 4 will provide for the first time a catalog of orbital solutions for three-body systems.
  Because astrometry is only sensitive to the motion in the plane of the sky, there is a degeneracy in the orbital solutions leading to two possible solutions for the mutual inclination in three-body systems.

  We propose a hierarchical Bayesian framework to circumvent this degeneracy and unveil the mutual inclination distribution in the population.
  We tested our methods on simulated datasets, and showed that it reliably recovered the injected mutual inclination distributions in various cases.
  The method is thus promising for the statistical analysis of Gaia DR4 three-body systems.
}

\keywords{astrometry -- planets and satellites: general -- binaries: general -- methods: statistical}

\maketitle
\nolinenumbers

\section{Introduction}
\label{sec:introduction}

Astrometric surveys such as Gaia have the potential to detect a variety of architectures of systems:
multiple star systems, multi-planetary systems, circumbinary planets, circumstellar planets in binary systems, etc.
Having a large sample of such systems will allow us to better constrain their formation and evolution.
The mutual inclination between orbits in multiple systems is of particular interest for the orbital dynamics.
For instance, systems with large initial mutual inclinations can undergo Von Zeipel-Lidov-Kozai
\citep[ZLK, see][]{vonzeipel_1910_lapplication, lidov_1962_evolution,kozai_1962_secular}
cycles, where the inner orbit's eccentricity can reach very high values,
and which can lead, together with tidal dissipation to the shrinking of the inner orbit
\citep[e.g.,][]{fabrycky_2007_shrinking}.
The orbital migration through ZLK effect and tides is expected to leave some
footprints in the systems architecture: short period inner orbit and mutual inclination close to 40\degr{} or 140\degr{} \citep{fabrycky_2007_shrinking}.

The mutual inclination distribution in three-body systems has been studied with several techniques.
For visual triple star systems, \citet{worley_1967_multiple} compared the apparent direction of motion in the sky of
both companions and found that apparently co-revolving orbits where more frequent than counter-revolving orbits,
thus showing a first evidence that orbits are more frequently aligned than anti-aligned.
This method has been used in subsequent works, with additional observational constraints \citep[e.g.,][]{sterzik_2002_relative,tokovinin_2017_orbit}.
In particular, \citet{tokovinin_2017_orbit} showed that the correlation between the apparent directions of motion in the sky of both companions depends on the separation of the outer orbit.
The most compact systems in their work (separation of 20-50~AU) are found to be strongly aligned,
while for separations of about 1000~AU the orbits orientations are uncorrelated.

For much more compact systems, \citet{borkovits_2016_comprehensive} studied eclipse timing variations (ETV) in eclipsing binaries observed by the \textit{Kepler} mission to detect and constrain the orbit of outer (non-eclipsing) stellar companions.
They detected outer companions with periods up to about 2000~d.
They could constrain the mutual inclination by modeling dynamical interactions in the triple star systems for 62 of these systems.
No system was found with a mutual inclination larger than 60\degr, which is coherent with the trend
observed in visual triples that the most compact systems seem to be preferentially aligned.
The distribution of mutual inclinations they obtained \citep[see Fig.~15 of]{borkovits_2016_comprehensive}
is bimodal, with a main mode at 0\degr{} and a secondary peak close to 40\degr,
which could be interpreted as an evidence for ZLK-migration.

In this context, Gaia astrometry could provide additional observational constraints,
in an intermediate regime of separation, and for lower mass companions.
With the up-coming Data Release 4 (DR4)\footnote{\url{https://www.cosmos.esa.int/web/gaia/dr4}},
the Gaia collaboration will publish 16'681 Non-Single-Star (NSS) Multiple Orbits candidates.
These NSS Multiple Orbits solutions include two-body orbits with additional acceleration and jerk terms,
as well as three-body solutions.
The exact number of actual three-body solutions is not yet known but could be enough to perform statistical analyses.

The astrometric detection method allows one to measure all orbital parameters,
including the inclination and longitude of the node of each planetary or stellar companion.
However, there remains a degeneracy between two possible configurations for each companion:
one cannot tell which node is the ascending node and which is the descending node of the orbit.
In multiple systems, this degeneracy leads to a degeneracy between two possible solutions
for the mutual inclination.

For individual systems, and without any prior knowledge of the inclination distribution,
the astrometric degeneracy can only be solved with complementary observational constraints (e.g., radial velocities).
However, in this letter we explore the possibility to circumvent the degeneracy and recover
the underlying distribution of mutual inclination by
analyzing in a hierarchical Bayesian
statistical framework a large sample of systems detected through
an astrometric survey such as Gaia.

In Sect.~\ref{sec:mutincdistrib} we present
the statistical framework we propose to constrain the mutual inclination distribution in a population of systems.
In Sect.~\ref{sec:sim} we generate several populations of systems with
various mutual inclination distributions and test the performance of our method in recovering
these distributions.
In Sect.~\ref{sec:discussion} we discuss our results.

\section{Uncovering the mutual inclination distribution}
\label{sec:mutincdistrib}

In this section we present a statistical framework to constrain the mutual inclination distribution of a sample of systems detected through astrometry.
We first describe the geometry of the problem and the fundamental degeneracy
encountered for individual systems (Sect.~\ref{sec:geom}).
We then estimate the marginal likelihood of the mutual inclination for individual systems (Sect.~\ref{sec:marginal}),
which we then use to explore the underlying mutual inclination distribution in the sample of systems (Sect.~\ref{sec:underlying}).

\subsection{Geometry of the problem}
\label{sec:geom}

\begin{figure}
  \centering
  \includegraphics[width=0.7\linewidth, trim=2mm 11mm 6mm 4mm, clip]{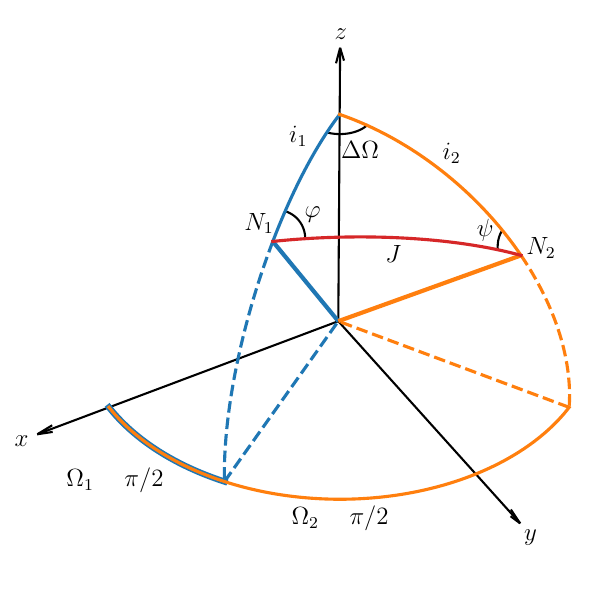}
  \caption{Geometry of the inclinations in a three-body system.
    The line of sight is along the $z$-axis, and the reference of longitudes is along the $x$-axis.
    The points $N_1$ and $N_2$ are the projections of the angular momentum of both companions' orbits on the unit sphere (north poles).}
  \label{fig:spherical_triangle}
\end{figure}

Figure~\ref{fig:spherical_triangle} illustrates the geometry of the problem.
Given the orbital elements of two companions orbiting the same star,
their mutual inclination $J$ is given by (spherical laws of cosines):
\begin{equation}
  \label{eq:cosJ}
  \cos J = \cos i_1 \cos i_2 + \sin i_1 \sin i_2 \cos \Delta\Omega,
\end{equation}
where $i_k$ are the inclinations of both companions with respect to the line of sight,
and $\Delta \Omega = \Omega_2 -\Omega_1$ is the difference of their longitudes of ascending nodes.

Using astrometry alone, and due to the projection of the orbits on the plane of the sky,
one cannot distinguish between an orbit with argument of periastron $\omega$ and longitude of node $\Omega$
and an orbit with $\omega' = \omega+\pi$ and $\Omega'=\Omega+\pi$.
Therefore, for two companions, one cannot distinguish between $\Delta\Omega$ and $\Delta\Omega + \pi$,
which thus provides two possible solutions for the mutual inclination (Eq.~(\ref{eq:cosJ})).
By convention, Gaia astrometric solutions always provide the solution for which
$\Omega\in[0,\pi[$.
        We denote this solution by $\tilde\Omega = \Omega \mod \pi$,
        and introduce
      $\Delta\tilde\Omega = \Delta\Omega \mod \pi$.

        \subsection{Marginal likelihood for the mutual inclination}
        \label{sec:marginal}

        We start from the maximum-likelihood (ML) estimate $\hat\theta$ of all the fitted parameters
        as well as the covariance matrix $C_\theta$ such that the likelihood can be approximated (in the vicinity of the ML solution) as:
        \begin{equation}
          \log\mathcal{L}(\theta) = \log p(y|\theta) \approx -\frac{1}{2} (\theta-\hat\theta)\T C_\theta^{-1}(\theta-\hat\theta) + c,
        \end{equation}
        where $y$ is the set of observations
        and $c$ is a constant with respect to the parameters $\theta$.
        We aim here at computing the marginal likelihood:
        \begin{equation}
          \mathcal{L}(J) = p(y|J) = \int p(y|\theta) p(\theta|J) \d\theta
        \end{equation}

        In Gaia DR4\footnote{\url{https://www.cosmos.esa.int/web/gaia/dr4}},
        the fitted parameters for three-body solutions include the Thiele-Innes parameters (see Appendix~\ref{app:orbpar}) for both companions,
        but the DR4 tables will also provide the corresponding Campbell elements (see Appendix~\ref{app:orbpar}),
        as well as the full covariance matrix for all parameters,
        evaluated at the ML solution.

        We thus start here from the Campbell elements, but introduce the additional change of variables from the two longitudes of the nodes $(\tilde\Omega_1, \tilde\Omega_2)$
        to $(\tilde\Omega_1, \Delta\tilde\Omega)$, and propagate the covariance matrix accordingly.

        We then introduce $\eta = (i_1, i_2, \Delta\tilde\Omega)$,
        and $\theta_\mathrm{other}$ all other parameters, such that
      $\theta = (\eta, \theta_\mathrm{other})$.
        We assume wide uninformative priors for $\theta_\mathrm{other}$ such that
        \begin{eqnarray}
          \log p(y|\eta) &=& \log \int p(y|\theta) p(\theta_\mathrm{other}) \d\theta_\mathrm{other}\nonumber\\
          &\approx& -\frac{1}{2} (\eta -\hat{\eta})\T C_\eta^{-1} (\eta -\hat{\eta}) + c',
          \label{eq:loglike_eta}
        \end{eqnarray}
        where $C_\eta$ is the submatrix of $C_\theta$ corresponding to the parameters $\eta$.
        When evaluating this expression, the difference $\Delta\tilde\Omega - \widehat{\Delta\tilde\Omega}$
        should be brought back to $]-\pi/2, \pi/2]$.

We now need to compute the integral over $\eta$:
\begin{equation}
  \label{eq:marginal_likelihood_eta}
  \mathcal{L}(J) = \int p(y|\eta) p(\eta|J) \d\eta,
\end{equation}
which requires to set the conditional prior $p(\eta|J)$.

We assume that given the mutual inclination between both companions,
there is no preferred direction on the sphere for the observer's line of sight.
We note that, all other things being equal, a pole-on orbit
-- with an inclination close to 0 or 180\degr{} --
has a factor $\sqrt{2}$ increase in signal-to-noise ratio (SNR)
compared to an edge-on orbit
-- with an inclination close to 90\degr{} --
\citep[see][appendix C]{arenou_2023_gaia}.
Therefore, our assumption is not fully satisfied for low amplitude signals,
close to the detection limit in terms of SNR.

With this approximation, we have a uniform probability distribution of
the pole $N_1$ over the sphere, and thus:
\begin{equation}
  p(i_1|J) = p(i_1) = \frac{\sin i_1}{2}.
\end{equation}
Once the mutual inclination and the pole $N_1$ of the first orbit are known,
the pole $N_2$ of the second orbit is restricted to a circle of radius $J$ around $N_1$
(see Fig.~\ref{fig:spherical_triangle}).
Again, in the approximation that there is no preferred direction on the sphere for the observer's line of sight,
the probability distribution of the angle $\varphi$ determining the position of $N_2$ on this circle is uniform:
\begin{equation}
  p(\varphi|J,i_1) = \frac{1}{2\pi}.
\end{equation}
We thus have
\begin{equation}
  p(i_1,\varphi|J) = \frac{\sin i_1}{4\pi}.
\end{equation}
Using spherical trigonometry relations (laws of sines and cosines), we could perform the
change of variables from $(i_1, \varphi)$ to $(i_1, i_2)$ and obtain:
\begin{equation}
  p(i_1, i_2 | J) = \frac{1}{4\pi |\sin \Delta\Omega|}= \frac{1}{4\pi \sin \Delta\tilde\Omega}.
\end{equation}
However, computing the marginal likelihood through the integral
of Eq.~(\ref{eq:marginal_likelihood_eta}) over $i_1,i_2$ is numerically challenging
because of the singularities of $p(i_1, i_2 | J)$ at $\Delta\Omega = 0,\pi$.

To avoid these singularities, we actually do not perform the change of variables from $(i_1,\varphi)$ to $(i_1,i_2)$,
and instead define $u_1 = \cos(i_1)$ and $\zeta = (J, u_1, \varphi)$ such that
we have a uniform prior:
\begin{equation}
  p(u_1, \varphi|J) = \frac{1}{4\pi}
\end{equation}
and thus
\begin{eqnarray}
  \label{eq:marginal_likelihood_final}
  p(\zeta|J) &=& \frac{\delta_J}{4\pi},\nonumber\\
  \mathcal{L}(J) &=& \int p\left(y|\eta(\zeta)\right) p(\zeta|J) \d\zeta\nonumber\\
  &=& \frac{1}{4\pi} \iint p\left(y|\eta(u_1, \varphi, J)\right) \d u_1 \d\varphi.
\end{eqnarray}
We then use Eqs.~(\ref{eq:loglike_eta}) and (\ref{eq:marginal_likelihood_final})
to compute the marginal likelihood $\mathcal{L}(J)$ on a grid of mutual inclination $(J_i)_{(i=1,\dots,n_J)}$.
The double integral of Eq.~(\ref{eq:marginal_likelihood_final}) is evaluated numerically by sampling
$u_1$ and $\varphi$ on a regular grid.

\subsection{Underlying mutual inclination distribution}
\label{sec:underlying}

Using the method described in Sect.~\ref{sec:marginal}, we are able to precompute the value
of the marginal likelihood $\mathcal{L}_s(J_i)$ on a grid of mutual inclinations $(J_i)_{(i=1,\dots,n_J)}$,
and for each system $s$ in our sample.

Let us now assume that the distribution of mutual inclination in the population
can be parameterized with a set of parameters $x$.
The reasoning here is general and can be applied to any parametric distribution $p(J|x)$ over $[0,\pi]$,
In the applications of Sect.~\ref{sec:sim},
we approximate the distribution by a Gaussian mixture model (GMM)
and the parameters $x$ are the amplitudes, locations, and scales of the (truncated) Gaussian components.
But similar methods could be applied to a distribution modeled with splines, or flat distributions over bins, etc.

For a given set of distribution parameters $x$, the likelihood of the full set of observations is:
\begin{equation}
  \label{eq:all_systems_likelihood}
  \mathcal{L}(x) = \prod_{s=1}^{n_\mathrm{system}} \int \mathcal{L}_s(J) p(J|x) \d J,
\end{equation}
which we approximate by a sum on the grid of precomputed $\mathcal{L}_s(J_i)$ values:
\begin{equation}
  \mathcal{L}(x) \approx \prod_{s=1}^{n_\mathrm{system}} \sum_{i=1}^{n_J} \mathcal{L}_s(J_i) p(J_i|x) \delta J.
\end{equation}
The log-likelihood is thus:
\begin{equation}
  \label{eq:loglike_x}
  \log \mathcal{L}(x) = \sum_{s=1}^{n_\mathrm{system}} \log\left( \sum_{i=1}^{n_J} \mathcal{L}_s(J_i) p(J_i|x) \right) + c'',
\end{equation}
where $c''$ is a constant with respect to the parameters $x$ that we can ignore since we are only interested in
likelihood ratios in the following.

We note that $\mathcal{L}_s(J_i)$ is a ($n_\mathrm{system}\times n_J$) matrix which can be precomputed,
while $p(J_i|x)$ is a vector of size $n_J$ which needs to be evaluated for each set of parameters $x$,
but only once for all systems.
We can then perform a MCMC exploration of the parameters $x$ using the likelihood of Eq.~(\ref{eq:loglike_x})
and assuming a prior distribution for $x$.

Using the same posterior samples, one can go back to individual systems and estimate the posterior
probability for the mutual inclination of each system, accounting for the observations from all systems
(see Appendix~\ref{app:postsingle} for the detailed calculations).

\section{Performance tests on simulated data}
\label{sec:sim}

In order to test the performance of our methods, we simulated 8 populations of systems assuming
different mutual inclination distributions.
Ideally, an end-to-end test would be to randomly sample all the orbital parameters for each fake system,
then to generate an astrometric timeseries with a realistic sampling and noise,
which would then be analyzed to search for the companions and estimate their orbital parameters.
Finally for systems for which both companions are reliably detected, their orbital parameters estimates and covariance matrix
would be used as input to our method.

We decided to simplify this procedure and directly generate fake estimates of the Thiele-Innes parameters of both companions.
This avoids generating and fitting fake measurements and also avoids having to generate the full set of orbital parameters,
since Thiele-Innes parameters do not depend on the eccentricity, period, or orbital phase.
We additionally assume the covariance matrix of the Thiele-Innes parameters to be diagonal (i.e., uncorrelated parameters).
Finally, the actual semi-major axes of the two orbits are not directly relevant to our study, since we are only interested in angles.
However, we need to specify how the uncertainties $\sigma_{TI}$ on the Thiele-Innes parameters compare with the semi-major axis $a$ (i.e., the SNR of the orbit).
We randomly sampled the SNR ($a/\sigma_{TI}$) of both orbits
assuming a log-uniform distribution of SNR between 3-30 for the inner orbit,
and between 5-50 for the outer orbit.

With these approximation, we need to generate the inclination $i_i$, argument of periastron $\omega_i$, and longitude of ascending node $\Omega_i$ for both companions in each system.
We first generate the mutual inclination $J$ according to the chosen distribution.
Then we sample the inclination and longitude of ascending node of the first companion, so that the pole $N_1$ of the
orbit is uniformly sampled on the sphere (i.e., $\cos(i_1)$ is sampled uniformly over $[-1,1]$ and $\Omega_1$ is sampled uniformly over $[0,2\pi[$).
We then uniformly sample the angle $\phi$ (see Fig.~\ref{fig:spherical_triangle}) over $[0,2\pi[$ to define the direction of the the pole $N_2$ of the orbit of the second companion.
We then derived its inclination $i_2$ and longitude of ascending node $\Omega_2$ using spherical trigonometry.
The arguments of periastron of both companions are uniformly sampled over $[0,2\pi[$ independently.

From these set of Campbell elements, we derive the normalized Thiele-Innes parameters for both companions in each fake system.
We then introduce Gaussian white noise (with standard deviation $\sigma_{TI}/a$ for each companion)
to generate the fake estimates of these parameters,
and then switch back to the Campbell elements.
We also propagate the covariance matrix of the Thiele-Innes elements to the Campbell elements.

With this procedure, we obtained sets of fake Campbell elements estimates on which we can directly apply our methods.
For each simulated population of systems, we first precomputed the marginal likelihood for each system on grid of mutual inclination $J$,
following Sect.~\ref{sec:marginal}.
We used a grid of 180 values of $J$, and computed the integral of Eq.~(\ref{eq:marginal_likelihood_final})
for each $J$ by sampling $u_1=\cos(i_1)$ and $\varphi$ on a grid of 180 values for both.

\begin{figure}
  \centering
  \includegraphics[width=0.9\linewidth, trim=2mm 2mm 2mm 2mm, clip]{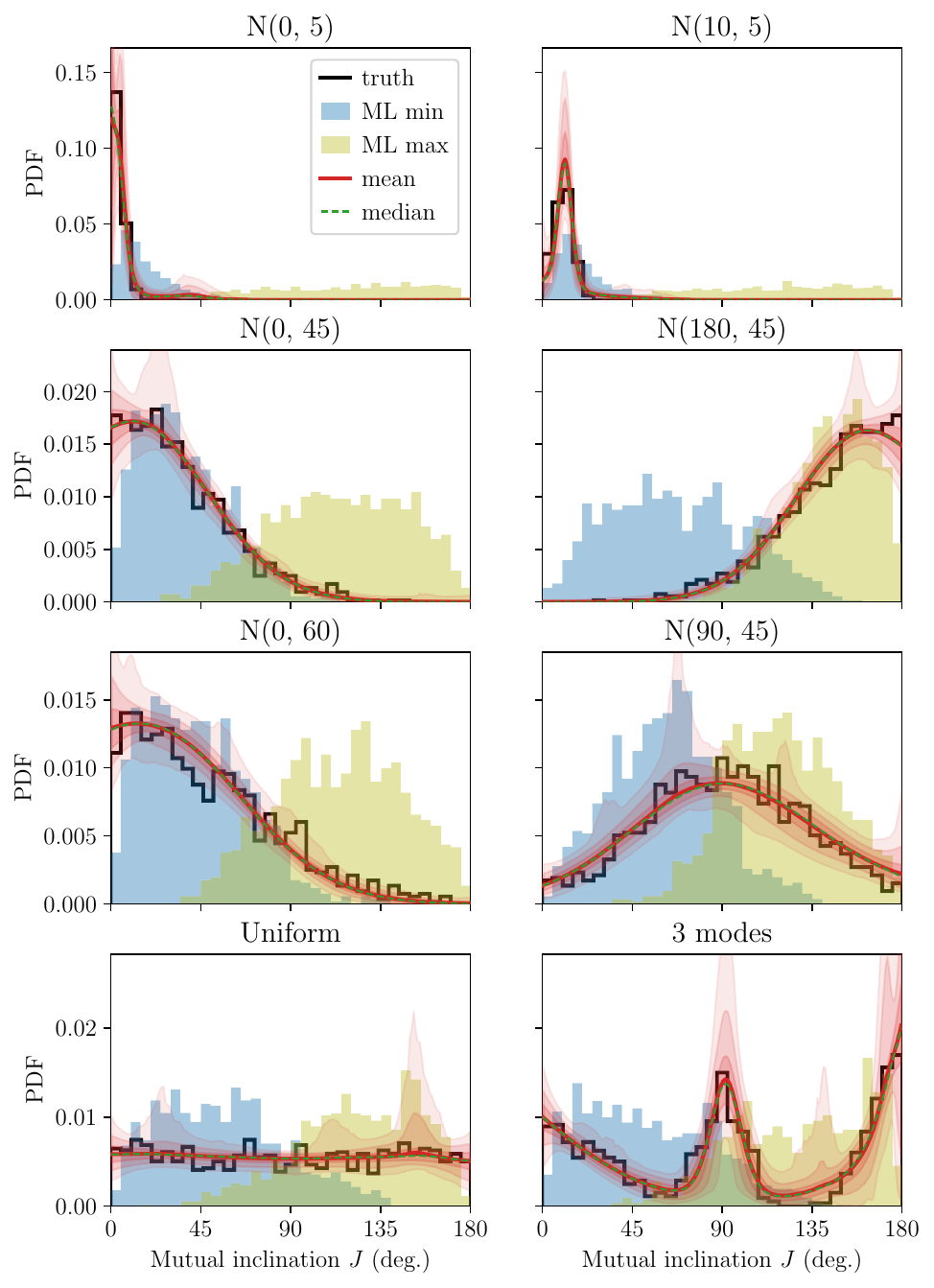}
  \caption{Mutual inclination distributions (PDFs) for the 8 simulated datasets of 1'000 systems each.
    The histogram of the true mutual inclinations (before simulating noise) is shown in \textit{black}.
    The histograms of the two ML solutions (minimum and maximum) for the mutual inclination of each system derived from the observed (noisy) Thiele-Innes parameters are shown in \textit{filled blue} and \textit{olive}.
    The mean of the posterior PDF is shown in \textit{red} as well as the 1,2, and 3-$\sigma$ intervals.
    The median of the posterior PDF is shown in \textit{dashed green}.
  }
  \label{fig:simus_pdf}
\end{figure}

We then modeled the mutual inclination distribution using a GMM with 1, 2, 3, or 4 components.
We followed the \textbf{baredSC}\footnote{\url{https://github.com/lldelisle/baredSC}} approach \citep{lopez_2022_baredsc} for the implementation of the GMM and explored the parameter space with the \textbf{samsam}\footnote{\url{https://gitlab.unige.ch/delisle/samsam}} MCMC package.
As in \citet{lopez_2022_baredsc},
we compute the evidence of the models with 1-4 components using importance sampling.
We assume a uniform prior for the number of components and combine all posterior samples, with any number of components.
Figure~\ref{fig:simus_pdf} shows the posterior for the probability density function (PDF) using these combined posteriors compared with the actual simulated distributions, for the 8 simulated datasets.
The mutual inclination distributions are well recovered in all simulated cases (normal distributions with various locations and scales, uniform, or multi-modal distribution).
We additionally show in Fig.~\ref{fig:simus_pdf} the two distributions of ML solutions for the mutual inclination (minimum and maximum solution for each system).
We see that these ML distributions are significantly different for the 8 simulations,
which gives an insight into how the mutual inclination distribution can be recovered from the data (see Appendix~\ref{app:observables} for more details).

\section{Discussion}
\label{sec:discussion}

We propose a method to uncover the mutual inclination distribution of three-body systems
detected with an astrometric survey such as Gaia.
Astrometry being only sensitive to the motion in the plane of the sky,
it has an intrinsic degeneracy which results in two solutions for the mutual inclination in three-body systems.
However, by analysing a sample of systems with a hierarchical Bayesian model,
we are able to circumvent this degeneracy and to
recover the mutual inclination distribution in the population.

As a side product, our method allows to derive the mutual inclination posterior for
individual systems (see Appendix~\ref{app:postsingle}).
This is roughly equivalent to re-analysing a system taking as prior the distribution found in the population.
While this could help resolving the degeneracy for a specific system,
it relies on the hypothesis that the system is representative of the population.

We tested our framework on simulated populations of systems with various mutual inclination distribution and were able to
reliably recover these distributions in all cases.
There are a few caveats in the method.
First, we assume that the likelihood can be approximated by a normal distribution over the Campbell elements
($i_1, i_2, \tilde\Omega_1, \tilde\Omega_2$).
For well constrained systems (high SNR), this is a good approximation,
but at lower SNR, this is probably less accurate,
due to non-linear correlations between the parameters.
Second, we assume that the observer direction is uniformly sampled on the sphere.
Again, this a good approximation at high SNR, because we don't expect specific alignment of the systems with
the Solar System direction.
However, a pole-on orbit has a factor $\sqrt{2}$ increase in SNR
compared to an edge-on orbit \citep[see][appendix C]{arenou_2023_gaia}.
Thus, at low amplitude, pole-on orbits are easier to detect,
and the observer direction is no more uniform on the sphere.
If the sample is dominated by systems were both companions are at low SNR,
this might lead to a bias favoring aligned and anti-aligned orbits with respect to mutual inclination of 90\degr.
If most analyzed systems present at least one high SNR companion (thus without a preferred orientation),
the bias should be low.

This method is promising in the context of the up-coming Gaia DR4 and its sample of multiple orbits solutions.
If a significant fraction of the 16'681 NSS Multiple Orbits candidates that will be published with DR4
are actual three-body systems, we will be able to derive the mutual inclination distribution in the global population,
or even in specific sub-populations.

\begin{acknowledgements}
  We thank the anonymous referee for their constructive feedback that helped to improve this manuscript.
  The author acknowledges the financial support of the Swiss National Science
  Foundation (SNSF), supported since May 2022 over the grant 200020\_205010.
  This work has been carried out within the framework of the National Centre of
  Competence in Research PlanetS supported by the SNSF under grants
  51NF40\_182901 and 51NF40\_205606.
\end{acknowledgements}

\bibliographystyle{aa}
\bibliography{biblio}

\begin{appendix}
  \nolinenumbers

  \section{Campbell and Thiele-Innes orbital elements}
  \label{app:orbpar}

  In this appendix we recall the definitions of the Campbell and Thiele-Innes orbital elements.
  The Campbell elements are the most widely used set of orbital parameters:
  period $P$, semi-major axis $a$, time of periastron passage $T_p$, eccentricity $e$, argument of periastron $\omega$, inclination with respect to the line of sight $i$, and longitude of ascending node $\Omega$.

  The Thiele-Innes parameters $(A,B,F,G)$ are more specific to astrometry and describe how the orbit projects onto the plane of the sky:
  \begin{eqnarray}
    A &=& a \left(\cos\omega \cos\Omega - \sin\omega \sin\Omega \cos i\right),\\
    B &=& a \left(\cos\omega \sin\Omega + \sin\omega \cos\Omega \cos i\right),\\
    F &=& a \left(-\sin\omega \cos\Omega - \cos\omega \sin\Omega \cos i\right),\\
    G &=& a \left(-\sin\omega \sin\Omega + \cos\omega \cos\Omega \cos i\right).
  \end{eqnarray}
  Denoting by $e$ the eccentricity and by $E$ the eccentric anomaly,
  the orbital motion in the plane of the sky is written as
  \begin{eqnarray}
    \delta &=& A \left(\cos E - e\right) + F \sqrt{1-e^2} \sin E,\\
    \alpha^* &=& B \left(\cos E - e\right) + G \sqrt{1-e^2} \sin E,
  \end{eqnarray}
  where $\delta$ (declination), and $\alpha^* = \alpha\cos\delta$ (modified right-ascension) are
  the local rectangular coordinates in the plane of the sky.

  \section{Mutual inclination of individual systems}
  \label{app:postsingle}

  With the MCMC approach of Sect.~\ref{sec:underlying}, we obtain
  posterior samples of the parameters $x$ of the mutual inclination distribution:
  $(x_k)_{(k=1,\dots,n_\mathrm{samp.})}$
  following $p(x|Y)$, where $Y$ is the set of all observations (from all three-body systems).

  Here we are interested in exploring the posterior probability for the mutual inclination of a specific system $s$ knowing all observations:
  $p(J_s | Y)$.
  This probability can be written as:
  \begin{equation}
    p(J_s | Y) = \int p(J_s | Y, x) p(x | Y) \d x.
  \end{equation}
  We assume that $p(J_s | Y, x) = p(J_s | y_s, x)$,
  i.e., that the observations of all the other systems (except $y_s$)
  do not bring any information about $J_s$ that is not already captured by
  the distribution parameters $x$.
  Thus
  \begin{eqnarray}
    p(J_s | Y) &=& \int p(J_s | y_s, x) p(x | Y) \d x\nonumber\\
    &=& \int \frac{p(y_s | J_s) p(J_s | x)}{p(y_s|x)} p(x | Y) \d x\nonumber\\
    &=& \int \frac{\mathcal{L}_s(J_s) p(J_s | x)}{\int \mathcal{L}_s(J) p(J | x) \d J} p(x | Y) \d x.
  \end{eqnarray}

  Finally, this integral can be approximated as a sum over all MCMC samples:
  \begin{eqnarray}
    p(J_s | Y) &\approx& \frac{1}{n_\mathrm{samp.}}\sum_{k=1}^{n_\mathrm{samp.}} \frac{\mathcal{L}_s(J_s) p(J_s | x_k)}{\int \mathcal{L}_s(J) p(J | x_k) \d J}.
  \end{eqnarray}
  Using the same grid of mutual inclinations as in Sect.~\ref{sec:marginal},
  we can estimate the posterior distribution for $J_s$ on the grid as:
  \begin{eqnarray}
    \label{eq:inc_posterior}
    p(J_i | Y) &\approx& \frac{\mathcal{L}_s(J_i)}{n_\mathrm{samp.}} \sum_{k=1}^{n_\mathrm{samp.}} \frac{p(J_i | x_k)}{\sum_{j=1}^{n_J} \mathcal{L}_s(J_j) p(J_j | x_k)\delta J}.
  \end{eqnarray}

  This approach is roughly equivalent to estimating the mutual inclination
  of a single system using a prior informed by all other systems.
  In such a case, the posterior distribution for the given system's mutual inclination would be:
  \begin{equation}
    \label{eq:indiv_inc_exploration}
    p(J_s | y_s) = \frac{p(y_s | J_s) p(J_s)}{\int p(y_s | J) p(J) \d J},
  \end{equation}
  where $p(J)$ is the prior for this analysis,
  and that could be determined from a previous analysis of all other systems.
  If the prior $p(J)$ is strongly constrained,
  even if the likelihood $p(y_s | J_s)$ is bimodal,
  the posterior $p(J_s | y_s)$ could favor one of the modes.

  Using the method of Eq.~(\ref{eq:inc_posterior}) offers two advantages
  compared to Eq.~(\ref{eq:indiv_inc_exploration}):
  it reuses the posteriors samples from the population without requiring a new exploration,
  and it naturally takes into account the uncertainty we have
  on the mutual inclination distribution in the population by integrating over the sampled
  parameters $x$.

  In both cases, the degeneracy resolving for an individual system
  strongly relies on the hypothesis that the considered system is representative of the global population.

  \section{Insights into the degeneracy resolving}
  \label{app:observables}

  \begin{figure}
    \centering
    \includegraphics[width=0.9\linewidth, trim=2mm 2mm 2mm 2mm, clip]{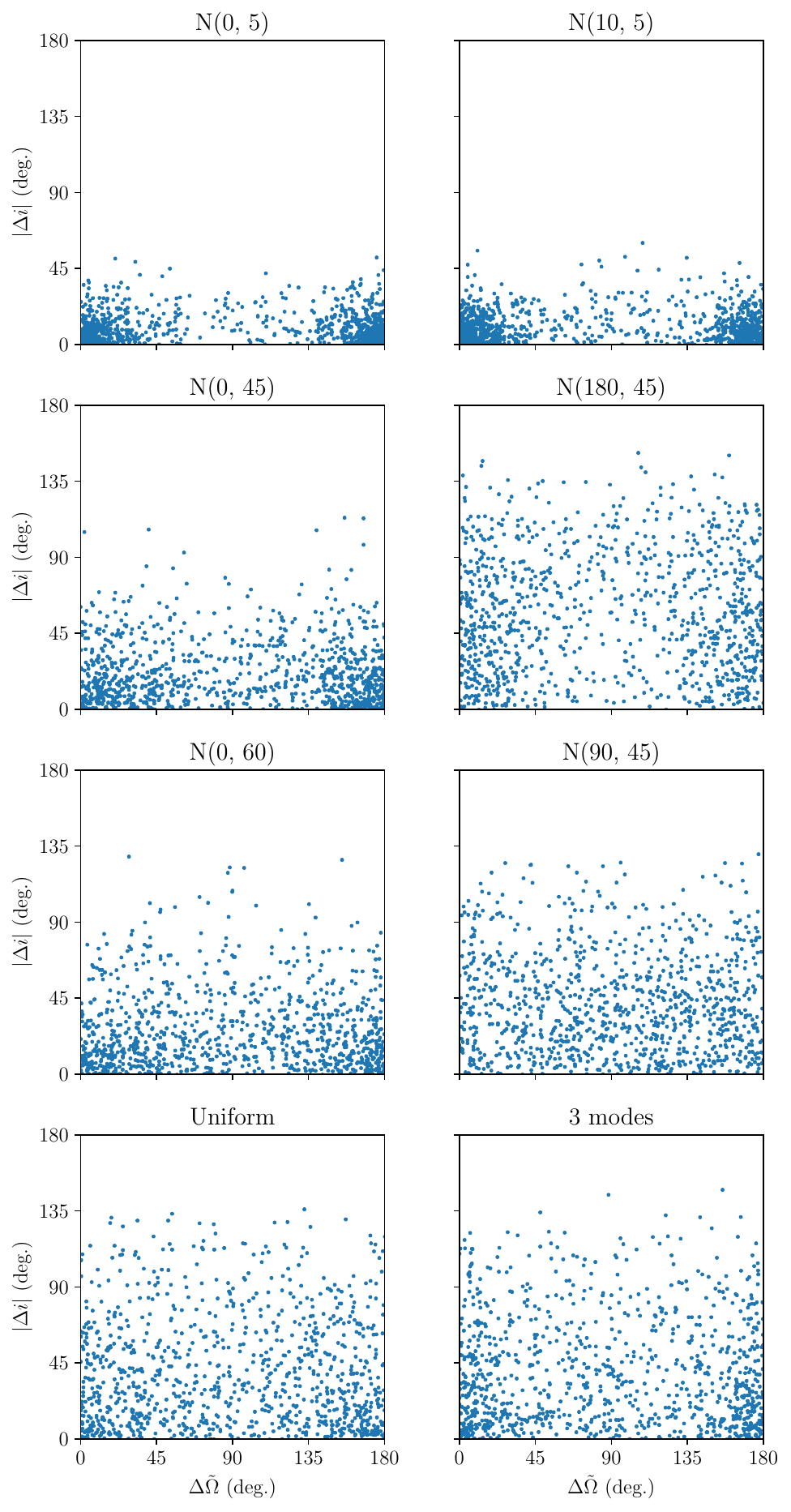}
    \caption{Scatter plots of the ML solutions for the 8 simulated datasets.
      We plot the difference of longitude of ascending node ($\Delta \tilde\Omega$) and absolute difference of inclinations ($|\Delta i|$) derived from the the observed (noisy) Thiele-Innes parameters.
    }
    \label{fig:simus_dW_di}
  \end{figure}

  In this appendix, we try and give a sense of how the method is able to resolve the degeneracy and recover the intrinsic mutual inclination distribution.
  Figure~\ref{fig:simus_dW_di} shows scatter plots of the difference of longitude of ascending node ($\Delta \tilde\Omega$) versus the absolute difference of inclinations ($|\Delta i|$) of the ML solutions for the 8 simulated datasets of Fig.~\ref{fig:simus_pdf}.
  The 8 simulated datasets have different signatures in this plane, which correspond to two observables.
  This gives an insight into how the population's mutual inclination distribution can be uncovered from the data.

  Let us first focus on the first set of simulations, which corresponds to quasi-coplanar systems ($\mathrm{N}(0, 5)$).
  In this case, both individual inclinations are always close to each other ($|\Delta i|$ is small)
  and $\Delta \tilde\Omega$ is close to 0 modulo $\pi$.
  In Fig.~\ref{fig:simus_dW_di}, all the systems accumulate in the two bottom corners.
  As expected for quasi-coplanar systems, one of the two degenerate solutions for the mutual inclination
  is found close to zero for all systems (ML~min, blue histogram of Fig.~\ref{fig:simus_pdf} \textit{top left}).
  The second solutions (ML~max, yellow histogram), however, spread over $[0, \pi]$, because of the random orientation of the systems with respect to the line of sight.
  We note that the ML~min solution is slightly shifted away from 0 and slightly wider than the true distribution of mutual inclination
  because of the simulated noise in the observations.

  If the true distribution of mutual inclinations was not strongly peaked at 0, it would be very unlikely to observe
  all systems in such a configuration that one of the two degenerate mutual inclination solutions is always close to zero.
  This would require each system to be oriented with respect to the observer such that both individual inclinations are close to each other ($|\Delta i|$ small) and $\Delta \tilde\Omega$ close to 0 modulo $\pi$.
  So the fact that the ML~min distribution is peaked close to zero in Fig.~\ref{fig:simus_pdf}
  is a strong evidence of preferentially aligned orbits.

  This hand-waving reasoning has its mathematical counterpart in Eq.~(\ref{eq:all_systems_likelihood}).
  Indeed, the likelihood of Eq.~(\ref{eq:all_systems_likelihood}) is the highest when
  the overlap is maximized between the assumed mutual inclination distribution $p(J|x)$ and the marginal likelihood $\mathcal{L}_s(J)$ for each system $s$.
  The marginal likelihood $\mathcal{L}_s(J)$ is typically bimodal with the two peaks corresponding to the ML~min and ML~max values.
  All systems have one of the modes at low mutual inclination, while the second modes spread over $[0, \pi]$.
  Thus, the likelihood is maximized when the set of parameters $x$ is chosen such
  that $p(J|x)$ is strongly peaked at low mutual inclinations.

  More generally, an accumulation for many systems of one of the two modes around a mutual inclination value $J$
  will favor (in terms of likelihood) sets of parameters $x$ such that the distribution $p(J|x)$
  has significant weight around $J$.
\end{appendix}
\end{document}